\documentclass[a4paper,notitlepage,11pt]{article}
\usepackage{cite}
\usepackage[numbers,sort&compress]{natbib}
\usepackage{pifont}
\usepackage{caption}
\usepackage{graphicx}
\usepackage[utf8]{inputenc}
\usepackage[left=1.50cm,
right=1.50cm,
top=2.00cm,
bottom=2.00cm
]{geometry}
\usepackage{amsmath}
\usepackage{amssymb}
\usepackage{lmodern}
\usepackage{braket}
\usepackage{bbold}
\usepackage{dcolumn}
\usepackage{bm}
\usepackage{multicol}
\usepackage{float}
\usepackage{comment}
\usepackage{authblk}
\fontfamily{lmss}

\begin{document}

\title{The irreversibility cost of purifying Szilard's engine:\\Is it possible to perform erasure using the quantum homogenizer?}

\author[1,2]{Maria Violaris}
\author[1]{Chiara Marletto}

\affil[1]{\small Clarendon Laboratory, University of Oxford, Parks Road, Oxford OX1 3PU, United Kingdom}
\affil[2]{Mathematical Institute, University of Oxford, Woodstock Road, Oxford OX2 6GG, United Kingdom}
\normalsize

\setcitestyle{square}

\maketitle

\begin{multicols}{2}
[
\begin{abstract}
Erasure is fundamental for information processing. It is also key in connecting information theory and thermodynamics, as it is a logically irreversible task. We provide a new angle on this connection, noting that there may be an additional cost to erasure, that is not captured by standard results such as Landauer's principle. To make this point we use a model of irreversibility based on Constructor Theory -- a recently proposed generalization of the quantum theory of computation. The model uses a machine called the ``quantum homogenizer", which has the ability to approximately realise the transformation of a qubit from any state to any other state and remain approximately unchanged, through overall entirely unitary interactions. We argue that when performing erasure via quantum homogenization there is an additional cost to performing the erasure step of the Szilard's engine, because it is more difficult to reliably produce pure states in a cycle than to produce mixed states. We also discuss the implications of this result for the cost of erasure in more general terms. \\
\end{abstract} 
]

\section{\label{sec:level1}Introduction}
\begin{figure*}[!htb]
\minipage{0.32\textwidth}
 \includegraphics[width=\linewidth]{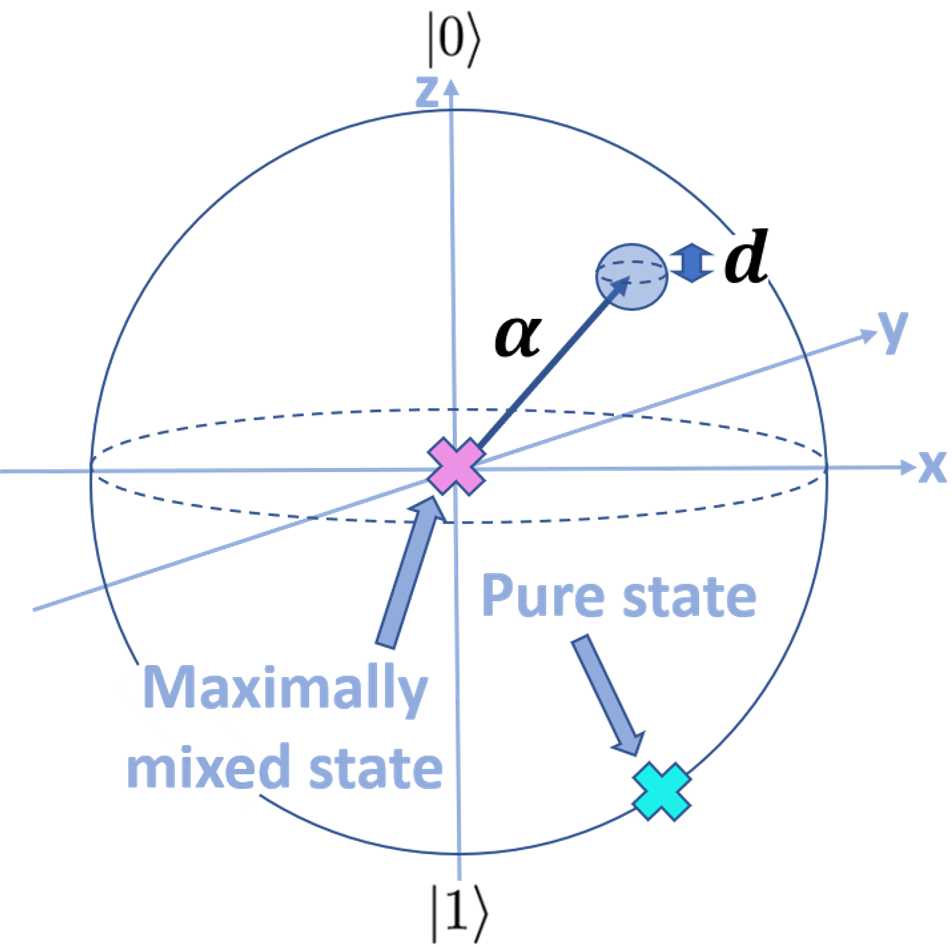}
 \caption{\label{Bloch sphere} The Bloch sphere and a Bloch vector of size $\alpha$, pointing to the original reservoir qubit state. Following homogenization, all the reservoir qubits, and the homogenized system qubit, are within some distance $d$ of the original reservoir qubit state, where $d$ can be made arbitrarily small as the size of the homogenizer is made arbitrarily large.}
\endminipage\hfill
\minipage{0.32\textwidth}
  \includegraphics[width=\linewidth]{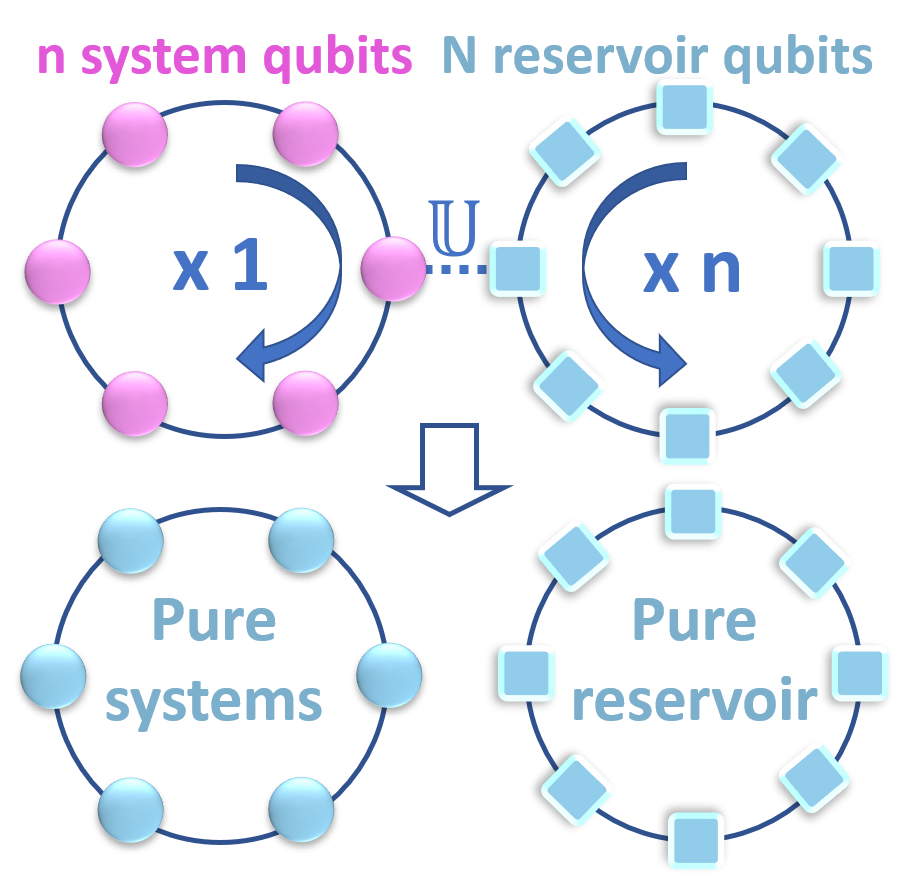}
  \caption{\label{quantum homogenizer} The quantum homogenizer, consisting of $N$ identical reservoir qubits, which each interact via the {\sl partial swap} with the same system qubit. Here we consider {\sl re-using} the machine to homogenize $n$ system qubits, as depicted above for a mixed-to-pure transformation, which may not be possible.} 
\endminipage\hfill
\minipage{0.32\textwidth}%
  \includegraphics[width=\linewidth]{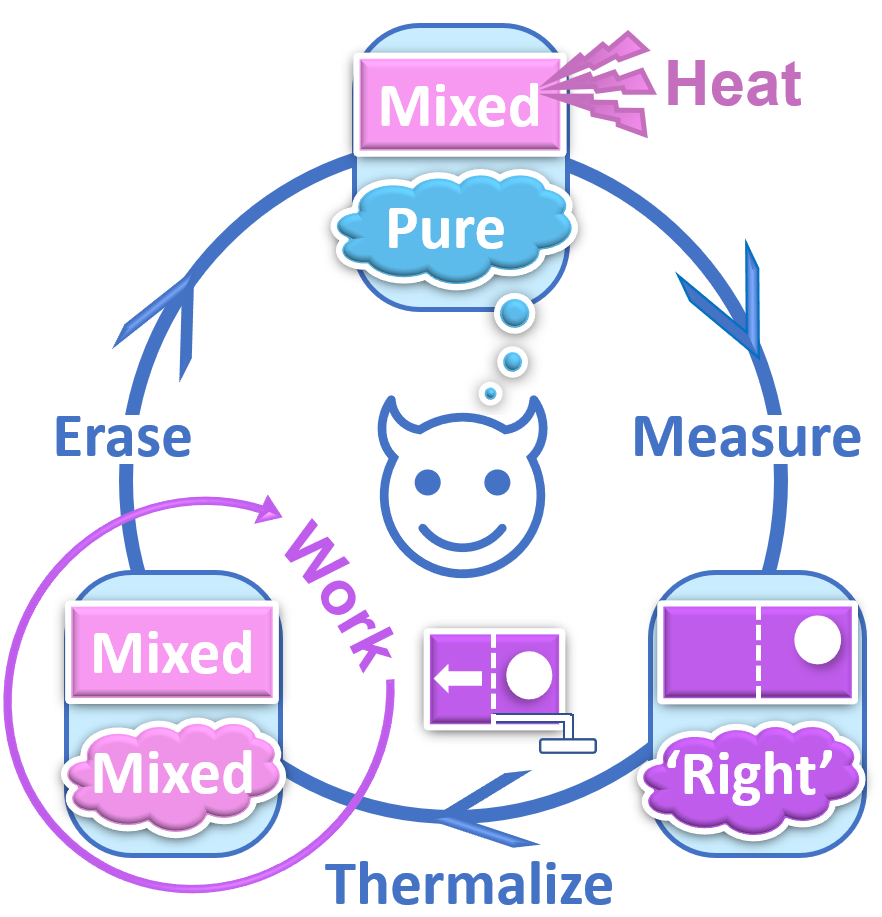}
  \caption{\label{szilard engine} Szilard's Engine. The ``demon" extracts work by measuring a particle's position in a box, inserting a sliding wall, then letting a weight be pulled up by the particle's pressure on the wall. The information stored about the particle must be erased to repeat the cycle, by transforming a mixed into a pure state.}
\endminipage
\end{figure*}

Erasure is essential for information processing and thermodynamics. The clearest realisation of the connection between erasure and thermodynamics is Bennett's application of Landauer's Principle to solve the Maxwell's Demon paradox \cite{Bennett_Landauer_1982}. The erasure of a demon's information about a system is essential for the demon to extract work in a cycle, and the erasure process has an irreducible entropy cost. This is most manifest in the case of the notorious {\sl Szilard engine}, the single-particle realisation of Maxwell's Demon (figure \ref{szilard engine}). For Szilard's engine to operate in a cycle, there must be a final step of memory erasure; and this is the step associated with an irreducible entropy cost -- see \cite{sep-information-entropy}. In the quantum model for Szilard's engine, memory erasure must transform a qubit from a mixed state to a pure state \cite{vedral_landauers_2000}. This is a non-unitary process, whence the irreducible entropy cost of this step. 

Here we consider whether there could be an additional cost to performing the erasure task via a particular method. This cost, as we shall illustrate, is associated to how well erasure can be performed using the so-called quantum homogenizer \cite{ziman_quantum_2001}, originally proposed as a unitary model to approximate the related non-unitary task of thermalization. 

In particular, we consider how far a machine can approximately realise the mixed-to-pure transformation needed for erasure, and approximately remain unchanged, so that it can work in a cycle. Various open-system dynamics can implement this transformation \cite{ng_limits_2015}, at the cost of causing changes to the environment in various ways. The quantum homogenizer can be used to implement this transformation, by partially swapping the mixed qubit to be erased with a qubit from a large reservoir of initially pure qubits. The protocol has been implemented experimentally for small numbers of qubits using NMR and photons, see e.g. \cite{violaris_nmr_2021, marletto_irreversibility_2022}.

In this paper we show that while a quantum homogenizer could in theory be constructed to perform a pure-to-mixed transformation to arbitrary accuracy -- crucially, while working {\sl in a cycle} -- such a homogenizer cannot be constructed to perform a mixed-to-pure transformation in a cycle. Specifically, in the mixed-to-pure case, the homogenizer deteriorates too quickly to perform the task indefinitely in a cycle. We conjecture that this captures an additional cost to operating Szilard's engine with the homogenizer, which goes beyond the entropy cost traditionally associated with erasure. We point out that this cost is not directly captured by a traditional analysis of entropy changes in terms of von Neumann entropy -- meaning that there may be more constraints to performing erasure than was previously thought. 

This cost is associated with  the so-called {\sl constructor-based irreversibility} \cite{marletto_irreversibility_2022}, which we build on in this paper. This type of irreversibility is exact and scale-independent, in contrast to many definitions of irreversibility using statistical mechanics. It has to do, informally, with the fact that a constructor, defined (following von Neumann \cite{Neumann_automata_1948}) as a machine that can be programmed to execute a task and retain the ability to perform it again, could be allowed for a task, but not for its transpose (i.e., the task where the inputs and outputs are switched). The fact that erasure is a task that may not necessarily be performable in a cycle, while its transpose is, suggests that erasure may have an additional link to irreversibility, beyond its traditional connection to the second law via Landauer's principle and Maxwell's Demon. This hints at the intriguing possibility that the second law can be updated in a stronger form, where it is exact and non-probabilistic, just like the law of conservation of energy is, \cite{marletto_work_2021}.

\subsection{Constructor-based irreversibility} \label{Constructors}

Our approach to explaining irreversibility is based on {\sl Constructor Theory} (CT) \cite{deutsch_constructor_2014, deutsch_constructor_2013, marletto_constructor_2015, marletto_constructor_2017}. In CT we define attributes as sets of states \cite{deutsch_constructor_2014}. The basic objects in the theory are \textbf{tasks}, defined as sets of ordered pairs of input and output attributes specifying a transformation on a physical system, or substrate.  One can then state the fundamental laws of physics in terms of whether tasks are {\bf possible} or {\bf impossible} and why, where possible and impossible are defined as follows. A {\bf constructor} for a task is a system capable of performing the task to arbitrarily high accuracy, while maintaining its ability to cause the transformation again. A task is possible ($T_{\checkmark}$) if it can be performed by a constructor, and impossible ($T_{\times}$) otherwise. For a possible task, there is no fundamental restriction on how closely a perfect constructor could be approximated by a sequence of (imperfect) actual machines.   

Generalizing this idea, one can conjecture the following ``constructor-based" irreversibility. Consider a possible task $T$, that sends some input attribute $\textbf{x}$ to an output attribute $\textbf{y}$. Its transpose $T^\sim $ is defined to have the inputs and outputs of $T$ switched, so it has input attribute $\textbf{y}$ and output attribute $\textbf{x}$. Then the task $T$ is possible, while its transpose $T^\sim $ need not be possible, even with time-reversal symmetric dynamics \cite{marletto_irreversibility_2022}:

\begin{equation} 
T_{\checkmark} = \{\textbf{x} \to \textbf{y}\},\hspace{0.5cm} T^{\sim}_{\times} = \{\textbf{y} \to \textbf{x}\}.
\end{equation}

The existence of a machine that can approach a perfect constructor for a task $T$ {\sl does not imply} that such a machine exists for the task $T^{\sim}$ in the reverse direction. This is because a machine approximately capable of performing a task in a cycle is not necessarily able to perform the transpose task in a cycle to the same degree of approximation, simply by having its dynamics reversed. The asymmetry holds even if the underlying dynamical laws are time-reversal symmetric, like those of quantum theory.

A model was recently proposed to illustrate this kind of irreversibility, based on unitary quantum theory \cite{marletto_irreversibility_2022}. Specifically, transforming a qubit from a pure state to a mixed state is possible, while transforming a qubit from a mixed to a pure state is not necessarily possible, even if this transformation is done via a series of unitary, time-reversal symmetric interactions. We first discuss this irreversibility in full generality within the formalism of quantum theory, and then demonstrate it when the task is performed by the quantum homogenizer \cite{ziman_quantum_2001}. 

\subsection{Quantum model for a constructor}

Here we discuss a general quantum model for a constructor, and provide an argument for the constructor-based irreversibility \cite{marletto_thesis_2016}. Consider the composite system of two quantum systems, $C$ and $S$, with total Hilbert space ${\cal H}={\cal H}_C\otimes {\cal H}_S$. Fix a unitary law of motion $U$ describing their interaction.

We can fix a task $T= \{{\bf x} \rightarrow {\bf y}\}$, where $\textbf{x}$ is the attribute associated with the density operator $\rho_{x}$ and $\textbf{y}$ is the attribute associated with the density operator $\rho_{y}$. For simplicity, from now on we will only consider attributes that do not change unless acted upon. In quantum theory, these attributes correspond to stationary states. Consider now the set of states of $C$ defined as follows:

\begin{equation}
 \Sigma^{(1)}_T=\{\rho_C \in {\cal H}_C \;:\;{\textrm{tr}}_{C}(U(\rho_C \otimes \rho_x)U^{\dagger}) = \rho_y\}.\\
\end{equation}

This is the set of states of $C$ with the property that, when $C$ is initialized in one of those states, and presented with the substrate $S$ in the state $\rho_x$ with the attribute ${\bf x}$, it delivers the substrate in a state with attribute ${\bf y}$. Note that in the final state $C$ and $S$ can be entangled. Hence, $C$ may no longer be able to cause the transformation again once it has performed it once.

We can check that $\Sigma^{(1)}_T$ is a convex set by considering how a convex combination of N density operators $\sum_{i=1}^{N}\lambda_i \rho_{C_i}, \sum_{i=1}^{N}\lambda_i = 1$, where $\rho_{C_i}$ are in $\Sigma^{(1)}_T$, cause a system state $\rho_x$ to evolve under the unitary $U$:\\

\begin{align}
\begin{split}
    & \textrm{tr}_C(U(\sum_{i=1}^{N}\lambda_i \rho_{C_i})\otimes \rho_x U^\dagger )\\
    & = \sum_{i=1}^{N}\lambda_i \textrm{tr}_C(U \rho_{C_i}\otimes \rho_x U^\dagger )\\
    & = \sum_{i=1}^{N}\lambda_i \rho_{y} = \rho_{y}.
\end{split}
\end{align}\\

Therefore the convex combination $\sum_{i=1}^{N}\lambda_i \rho_{C_i}$ is also in the set $\Sigma^{(1)}_T$, meaning that $\Sigma^{(1)}_T$ is a convex set.\\

Now let us proceed with a recursive definition of a family of sets: 

\begin{equation}
    \Sigma^{(n)}_T\subseteq \Sigma^{(n-1)}_T
\end{equation}

\noindent for all $n > 1$, with the property that $\forall \rho \in \Sigma_T^{(n)}$, setting $\rho_{out}=U(\rho\otimes \rho_x)U^{\dagger}$, we have that $\textrm{tr}_S\{\rho_{out}\} \in \Sigma_T^{n-1}$ and $\textrm{tr}_C\{\rho_{out}\} = \rho_y$.  

\begin{figure}[H]
\includegraphics[width=\columnwidth]{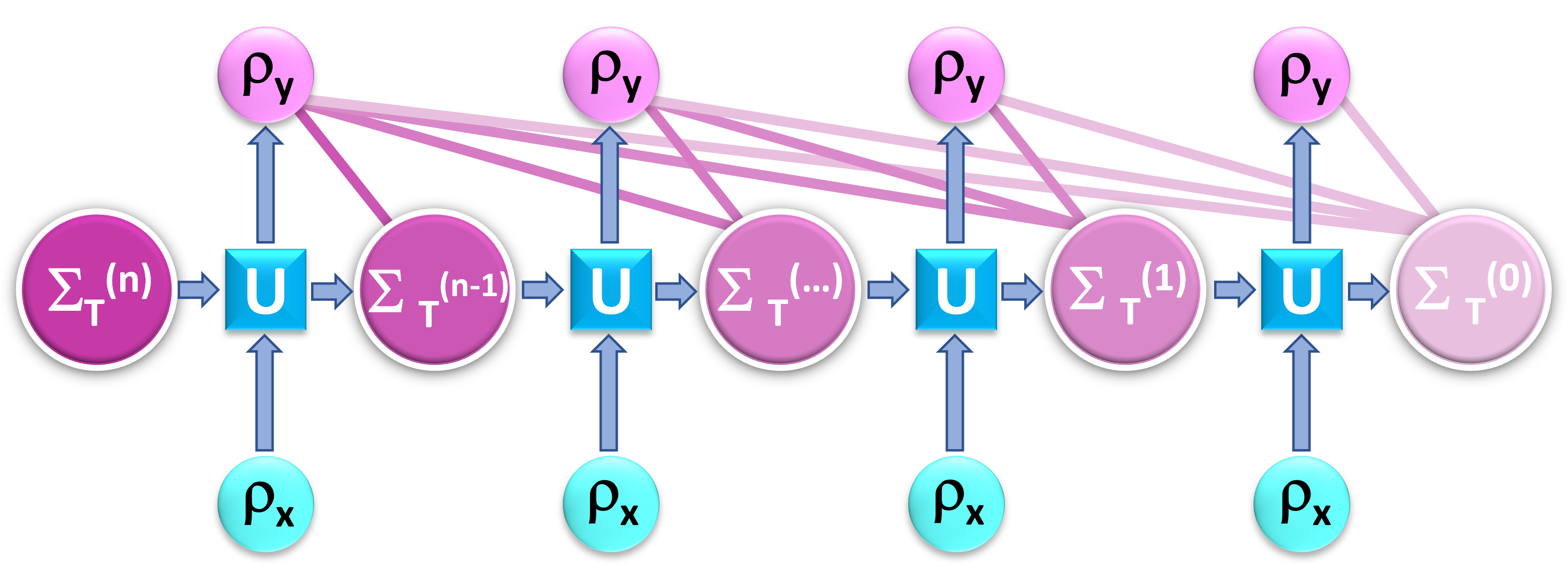}
\caption{\label{constructor_chain} Every time a machine in the set $\Sigma_T^{(i)}$ is evolved using unitary $U$ to transform a system from $\rho_x$ to $\rho_y$, it is left in a state in the set $\Sigma_T^{(i-1)}$. The lines between each $\rho_y$ and $\Sigma_T^{(i)}$ indicate that they may be entangled.}
\end{figure}

The set $\Sigma^{(n)}_T$ is the set of states of $C$ that can perform the task $T$ consecutively $n$ times, as depicted in Figure \ref{constructor_chain}.

After $n$ times $C$ may lose its ability to cause the task once again. A constructor is a special kind of environment which keeps this ability indefinitely. As such, it must be defined as the limit point of the sequence of sets $\{\Sigma^{(n)}_T\}_n$. 

A necessary condition for $C$ to be a {\sl constructor} for the task $T$ under the law of motion $U$ is that the set
\begin{equation}
    \Sigma_{C_T}\doteq \lim_{n\rightarrow \infty} \{\Sigma^{(n)}_T\}_n
\end{equation}
exists and it is non-empty. That the task $T$ is possible implies, in quantum theory, that there exists a non-empty set $\Sigma_{C_T}$ with the above properties. 

\subsubsection{The limit of a constructor}

A non-empty set $\Sigma_{C_T}$ defines a perfect constructor for a task: one which is able to perform the task arbitrarily well, an arbitrary number of times. In practice, if this machine is physically realized, it will not be a perfect constructor due to its inevitable deterioration from interactions with its environment. However, the existence of the non-empty set $\Sigma_{C_T}$ indicates that there is no limit to how well a perfect constructor for task $T$ can be approximated by a physical machine. 

For example, a CNOT gate and a qubit in the $\ket{1}$ state together form a  constructor for the task of transforming a state $\ket{0}$ to a state $\ket{1}$. This can be done by re-using the qubit in the $\ket{1}$ state as the control qubit with any number of target qubits in the state $\ket{0}$. However, physical CNOT gates cannot be implemented with no errors, and qubits cannot be prepared in the $\ket{1}$ state without errors. The transformation $\ket{0}$ to $\ket{1}$ is possible because there is no limit to how far the errors can be reduced in the CNOT implementation and $\ket{1}$ state preparation.  

\subsubsection{Constructor for transpose task}

We shall now argue that under quantum theory's time-reversal symmetric laws, the 
existence of a non-empty constructor set $\Sigma_{C_T}$ does not imply that a non-empty constructor set for the transpose task $T^{\sim}$,
$\Sigma_{C_{T^{\sim}}}$, must also exist. This statement holds even under the time-reversal $U^{\dagger}$ of the dynamical law. 

Consider again the task $T= \{{\bf x} \rightarrow {\bf y}\}$. Assume now that, according to time-reversal symmetry, all unitaries and their transposes are allowed. Assume also that $T$ is possible, so that the set $\Sigma_{C_T}$ exists and is non-empty under a given unitary $U$. This fact does not imply that the set $\Sigma_{C_{T^{\sim}}}$ must also be non-empty under $U^{\dagger}$, the time-reversal of $U$. For the inverse evolution $U^{\dagger}$ applied to $C$ prepared in $\Sigma_{C_T}$ and to the substrate $S$ initialized in the attribute ${\bf y}$, does not necessarily retrieve the substrate in the attribute ${\bf x}$. Hence we have the key result that $C$ initialized in the attribute $\Sigma_{C_T}$ is not necessarily a constructor for the task $T^{\sim}$ under the inverse unitary $U^\dag$.

This result can be understood as follows. If it is possible to perform the task $T= \{{\bf x} \rightarrow {\bf y}\}$, then in particular it must be possible to perform the task $T$ consecutively on $n$ systems for arbitrary $n$ -- meaning that the set $\Sigma_T^{(n)}$ is non-empty. One might then think that it must be possible to perform $T^{\sim}= \{{\bf y} \rightarrow {\bf x}\}$ consecutively on $n$ systems for arbitrary $n$ by reversing the unitary interaction -- meaning that as a result, the set $\Sigma_{T^{\sim}}^{(n)}$ is also non-empty. This is false. As depicted in Figure \ref{constructor_chain}, there can be entanglement (or more generally, correlations) between the system and machine following their interaction via $U$. Therefore to dynamically reverse the process in Figure \ref{constructor_chain} to convert $n$ systems in the state $\rho_y$ back to $n$ systems in the state $\rho_x$ using $U^\dag$, we need to prepare a very special initial state on the machine subspace. This initial state must not only have a reduced state in $\Sigma_T^{(0)}$ (the final state of a machine after performing the forwards task $T$ $n$ times), but also be entangled in a particular way with each of the $n$ systems to be transformed (all the systems $\rho_y$ shown in the top row of Figure \ref{constructor_chain}). 

However for the task $T^{\sim}= \{{\bf y} \rightarrow {\bf x}\}$ to be possible, we require that \textit{any} system having the attribute $\textbf{y}$ interacting with any machine state in $\Sigma_{C_{T^\sim}}$ will be transformed to a state having the attribute $\textbf{x}$, independently of whether the system is initially entangled with the machine or not. In other words, the transformation should work for any system whose reduced state is initially $\rho_y$, whether or not it is initially entangled with the machine, but this is not the case. If we applied the inverse unitary between the machine in Figure \ref{constructor_chain} with a system which was in the correct reduced state $\rho_y$ but not entangled with the machine in the correct way, then the inverse unitary would not in general transform the system to have the reduced state $\rho_x$. For let us denote by $U(\rho_C \otimes \rho_x)U^{\dagger} = \rho_{out}$ the global state of the machine and system after the interaction $U$. We know that, by unitarity, $U^{\dagger}\rho_{out}U = \rho_C \otimes \rho_x$. However, in general $U^{\dagger}(\rho_C \otimes \rho_y)U \neq \rho_C \otimes \rho_x$. Unlike the particular state $\rho_{out}$, a generic joint system and machine state, with reduced density operators $\rho_y$ on the system and $\rho_C$ on the machine, may not have the right correlations to allow $U^\dag$ to lead the joint state to $\rho_C\otimes\rho_x$. Hence while $\rho_C$ in $\Sigma_{C_T}$ may well be a state that represents a constructor for $T$, its existence does not imply that a constructor should exist for the transpose task $T^{\sim}$, even when dynamical laws are unitarily reversible. 

This argument shows that irreversibility in Constructor Theory, stating that a task is possible but its transpose need not be, is compatible with time-reversal symmetric laws, in an exact way. This is because of the fundamental difference between a task being possible and a dynamical law being allowed. We will now look at a particular implementation of this constructor-based irreversibility within an eraser, implemented in a qubit model for homogenization. 

\subsection{The quantum homogenizer}

The concept of re-using the same physical entity to perform a task occurs frequently in thermodynamics. When an item is put in a refrigerator to be cooled, it is expected that the refrigerator will retain the ability to perform the same task with a new item. Eventually, after cooling many items, any physical refrigerator breaks down and loses the ability to perform the task. Here, the physical entity is the quantum homogenizer, a set of initially identical reservoir qubits. The new input with each use is a fresh system qubit, and the task is to {\sl homogenize} the state of that system qubit, such that it becomes close to the state of the original reservoir qubits. An effective homogenizer will also remain itself almost unchanged by the interaction, retaining the ability to perform the task again. We will compare the effectiveness and robustness of this machine in two scenarios: each input system qubit is pure, and the original reservoir qubits are maximally mixed (``pure-to-mixed"); and each input system qubit is maximally mixed, and the original reservoir qubits are pure (``mixed-to-pure"). 

The quantum homogenizer is a machine consisting of $N$ identical {\sl reservoir} qubits (figure \ref{quantum homogenizer}). These each interact, one by one, with the {\sl system} qubit (the qubit whose state is to be transformed) via a unitary {\sl partial swap}: 

\begin{equation} \label{partial swap}
U = \text{cos}\eta \mathbb{1} + i\text{sin}\eta \mathbb{S}.
\end{equation}

The partial swap is a combination of the identity, $\mathbb{1}$ and swap operation, $\mathbb{S}$, weighted by the coupling strength parameter $\eta$. It has been shown that if the system qubit interacts with $N$ reservoir qubits via the partial swap, then as $N \to \infty$, the system qubit state converges to the original state of the reservoir qubits, for {\sl any} coupling strength $\eta \neq 0$ \cite{ziman_quantum_2001}. Furthermore, all of the reservoir qubits after the interaction are within some distance $d$ of their original state (see figure \ref{Bloch sphere}), which can be made arbitrarily small as coupling strength $\eta \to 0$. In the limit of the best possible homogenization, any system qubit $\rho$ is sent to the reservoir qubit state $\xi$, with all the reservoir qubits remaining unchanged:

\begin{equation}
    U^{\dagger}_N...U^{\dagger}_1 (\rho\otimes \xi^{\otimes N})U_1 ...U_N\approx\xi^{\otimes N+1}
\end{equation}

where $U_k := U \otimes (\otimes_{j\neq k} \mathbb{1}_j)$ denotes the interaction between the system qubit and the $k^{\text{th}}$ reservoir qubit. The information about the original system qubit state is seemingly erased, despite all the interactions being unitary and thus information-preserving. The information has actually become stored in the infinitesimal entanglement between infinitely many reservoir qubits, which sums to a finite value \cite{ziman_quantum_2001}.

{\bf Application to Szilard's engine.} Elaborating on recent studies, we show how the task of transforming a pure state to a mixed state is constructor-theoretically {\sl possible} using the quantum homogenizer, while the transpose task (mixed-to-pure) need not be \cite{marletto_irreversibility_2022}. The latter is the key task in the erasing step for a Szilard engine, inviting the question: does erasing a mixed state to a pure state in a cycle have an entropic cost {\sl in addition} to Landauer erasure's cost, due to the task's impossibility using the homogenizer? The analysis shown in this paper provides a basis from which to investigate a potential additional cost of erasure.

\section{Results} \label{Recurrence}

\subsection{Relative deterioration} \label{RelDet}

Here we show that in the weak coupling limit ($\eta \ll 1$), the quantum homogenizer can only be used as a constructor for the pure-to-mixed task and not the mixed-to-pure task. To conclude whether or not a task can be performed to arbitrary accuracy in a cycle by the homogenizer, one can consider the evolution of a quantity called the {\bf relative deterioration} \cite{marletto_irreversibility_2022}. This accounts for both the {\bf error} in performing the task, to quantify the potential for arbitrary accuracy, and the {\bf robustness} of the machine with multiple iterations, to quantify the potential for performing the task again. 

The accuracy in performing the task can be quantified as the fidelity of the output system qubit with the original state of the reservoir qubits. If the fidelity is close to 1, then the homogenization has been performed to high accuracy. Therefore one can define the error $\epsilon^n_N$ as this quantity subtracted from 1:

\begin{equation} 
\epsilon^n_N = 1 - F(\rho^n_N, \xi^0_j),
\end{equation}

where $n$ and $N$ are the total number of iterations and reservoir qubits respectively. Similarly, the robustness of the reservoir as it is used multiple times is the fidelity of the final reservoir state with the original reservoir state: 

\begin{equation} \label{robustness}
\delta^n_N = F(\xi^n_{N_{tot}}, \xi^{0 \otimes N}_j).
\end{equation}

The robustness $\delta^n_N$ defined in this way is 1 if the reservoir is unchanged after homogenization. The effects of error and robustness on performance of the task can be summarised using their ratio, namely relative deterioration $R^n_N$: 

\begin{equation} 
R^n_N = \frac{\epsilon^n_N}{\delta^n_N} = \frac{1 - F(\rho^n_N, \xi^0_j)}{F(\xi^n_{N_{tot}}, \xi^{0 \otimes N}_j)}.
\end{equation}

Relative deterioration scales with error and robustness such that in the limit of large number of iterations and large number of reservoir qubits (the condition for best accuracy), then there are two cases: 

\begin{equation} \label{Possibility definitions}
\lim_{n,N\to\infty} R^n_N = \begin{cases} 0 & \text{Task is possible,} \\
\infty & \text{Task need not be possible.}
\end{cases}
\end{equation}

If $R^n_N \to 0$ in the limit $n,N\to\infty$ then the deterioration with number of iterations increases slower than the error decreases with reservoir size. This means the task {\sl can} be performed to arbitrary accuracy in a cycle, so a {\bf constructor exists} for this task. If $R^n_N$ tends to $\infty$ in the limit $n,N\to\infty$ then deterioration with number of iterations increases faster than error decreases with reservoir size. This means the task {\sl cannot} be performed to arbitrary accuracy in a cycle by the homogenizer. Therefore, the homogenizer is {\bf not a constructor} for this task. 

The fidelity in the robustness expression is difficult to calculate analytically. Instead, the entire reservoir state can be approximated as the tensor product of the individual reservoir qubit states,  

\begin{equation} 
\xi^n_{N_{tot}} \approx \xi^n_1 \otimes ... \otimes \xi^n_N, 
\end{equation} 

where each $\xi^n_j = \text{Tr}_1(U(\rho^n_{j-1} \otimes \xi^{n-1}_j)U^\dag)$ is the local state of the $j^{\text{th}}$ reservoir qubit after $n$ iterations. Physically, the approximation amounts to neglecting the entanglement between the reservoir qubits. This is justified in the weak coupling limit, since inter-qubit entanglement becomes negligible for low $\eta$ \cite{ziman_quantum_2001}. 

Then the expression for the denominator of the relative deterioration, which measures the robustness of the machine, becomes: 

\begin{equation} \label{Fidelity product}
\begin{split}
F(\xi^n_{N_{tot}}, \xi^{0 \otimes N}_j)
& \approx F(\xi^n_1 \otimes ... \otimes \xi^n_N, \xi^{0 \otimes N}_j)\\
& = F(\xi^n_1, \xi^0_j)...F(\xi^n_N, \xi^0_j).
\end{split}
\end{equation} 

The derivations and final expressions for system and reservoir states, error, robustness and relative deterioration for the pure-to-mixed and mixed-to-pure transformations can be found in the Appendix. 

\begin{figure*}[!htb]
\minipage{\columnwidth}
\includegraphics[width=\columnwidth]{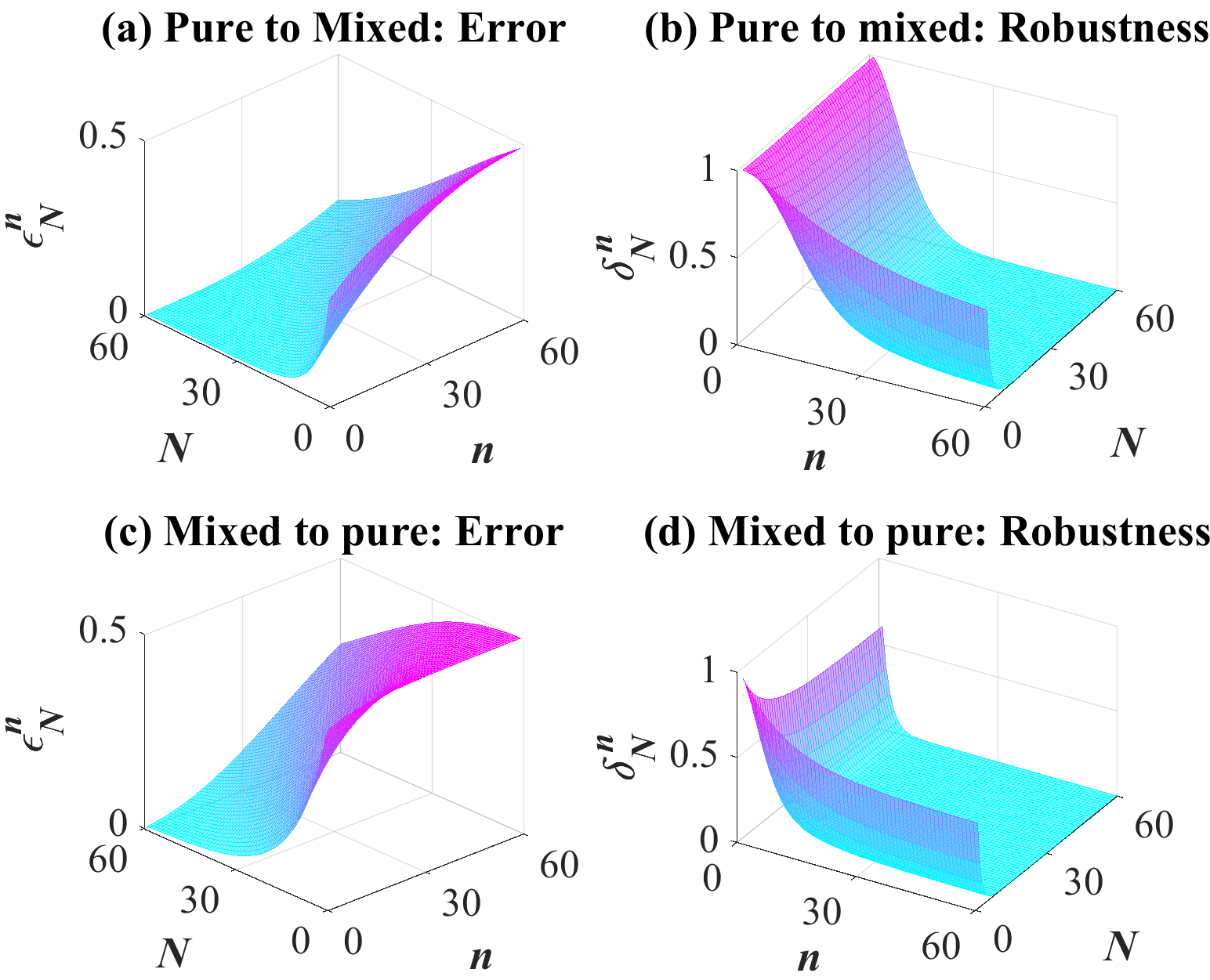}
\caption{\label{error and robustness 0.1} Evolution of error and robustness for a reservoir of size $N$ used $n$ times, with coupling strength $\eta = 0.5$.}
\endminipage\hfill
 \minipage{\columnwidth}
\includegraphics[width=\columnwidth]{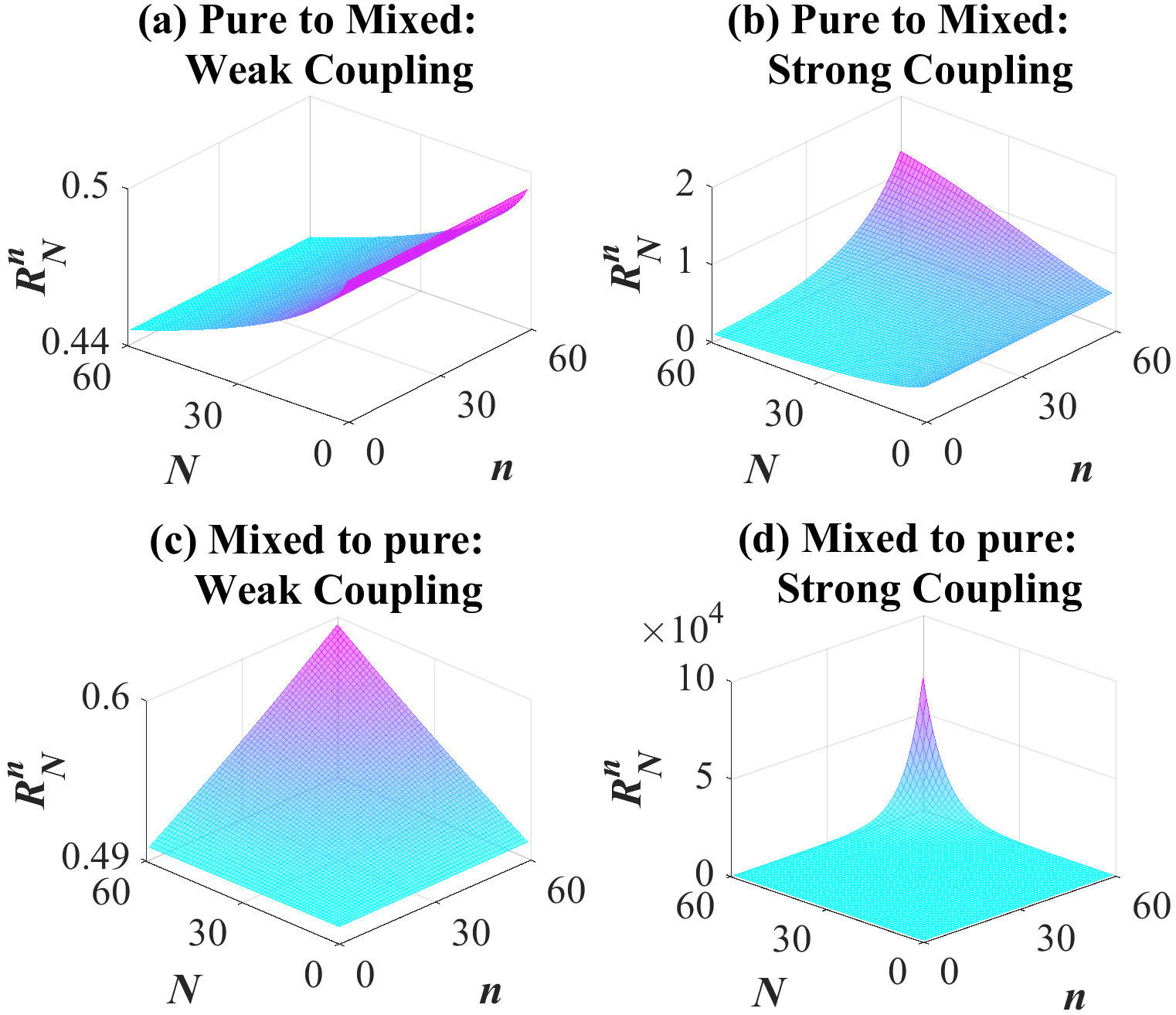}
\caption{\label{relative deterioration}Surfaces of relative deterioration for a reservoir of size $N$ used $n$ times, with $\eta = 0.01$ for weak coupling and $\eta = 0.1$ for strong coupling.}
\endminipage
\end{figure*}

\textbf{Patterns in relative deterioration.} The error and robustness for $N = n =$ 60, $\eta =$ 0.5 are shown in figure \ref{error and robustness 0.1}. 

Error decreases with increasing reservoir size, and increases with number of iterations, as expected. Conversely, robustness decreases with number of iterations as the reservoir deteriorates from its initial state with repeated use. The robustness also decreases with increasing reservoir size, as more reservoir qubits deteriorate from their initial state. The error for the mixed-to-pure case increases more with $n$ than for the pure-to-mixed case, and robustness for the mixed-to-pure case decreases more quickly and sharply with $n$ than for the pure-to-mixed case. These patterns indicate that the pure-to-mixed task can be performed {\sl more reliably} than the mixed-to-pure task. 

The relative deterioration reveals an asymmetry in the possibility of the two tasks. For a weak coupling of $\eta$ = 0.01, the surfaces of relative deterioration for the two directions of the task are shown in figures \ref{relative deterioration}(a) and (c). For large $N$ and $n$, $R^n_N \to 0$ for the pure-to-mixed case and $R^n_N \to \infty$ for the mixed-to-pure case. Using equation \ref{Possibility definitions}, the pure-to-mixed task is possible while the mixed-to-pure task is not possible when using the quantum homogenizer. 

\begin{figure}[H]
\includegraphics[width=\columnwidth]{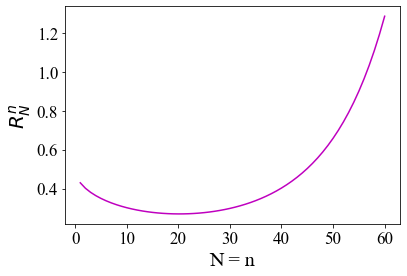}
\caption{\label{large N rel det 0.1} Approximate relative deterioration for a pure-to-mixed homogenizer used $N=n$ times, with coupling strength $\eta = 0.1$.}
\end{figure}

The second column of figure \ref{relative deterioration} shows that for the strong coupling regime, the relative deterioration remains significantly higher in the mixed-to-pure case than the pure-to-mixed case. However, the relative deterioration for large $n$ and $N$ is increasing and not tending towards 0 for the pure-to-mixed transformation. As can be seen more clearly in Figure \ref{large N rel det 0.1}, the relative deterioration decreases until around $N=n=20$, then begins increasing for larger $N=n$. 

This remains consistent with the pure-to-mixed transformation being possible using the homogenizer, since it need not be possible for all coupling strengths in order to be a possible task. It makes sense that the homogenizer has greatest reusability for small coupling strengths, as it is in the weak-coupling limit that the reservoir qubits are left almost unchanged after interacting with the system qubit. Additionally, in the strong coupling regime there is significant entanglement between the reservoir qubits that has been neglected due to the approximation in equation \ref{Fidelity product}. This has increasing impact as $N=n$ increases, demonstrated by the analysis of von Neumann entropy in section \ref{Von Neumann entropy}. This entanglement must be accounted for to determine conclusively how the relative deterioration behaves in the large $n$ and $N$ limit for strong coupling. 

\subsection{Von Neumann entropy} \label{Von Neumann entropy}
 
Now we will explain why von Neumann entropy is insufficient to capture the constructor-based irreversibility shown by the relative deterioration, and hence the additional cost of erasing mixed states to pure states. In any unitary evolution, the overall entropy must remain constant \cite{nielsen_quantum_2010}. In the pure-to-mixed task, the system qubit entropy increases by 1, so the entropy of the reservoir qubits must decrease by 1. The entropy changes occur in the opposite direction (but same magnitude) for the mixed-to-pure task. After each homogenization, the total of the reservoir entropies from the two tasks should remain constant, since one reservoir has decreased in entropy by the same amount that the other increased in entropy. The changes in entropy are symmetrical for the pure-to-mixed and mixed-to-pure cases, and do not indicate an entropic cost beyond what is expected from Landauer's principle. Therefore von Neumann entropy is insufficient to explain the asymmetry found in relative deterioration in the weak coupling regime.

We demonstrate this by considering the total entropy of the individual reservoir qubits from both tasks, $S_{tot}$. If the changes in entropy for the two tasks are equal and opposite, then the total entropy should remain constant as the homogenizer is used multiple times. This holds providing the entanglement between the reservoir qubits is negligible, because entanglement between qubits contributes negative entropy \cite{rio_thermodynamic_2011}. Hence, $S_{tot}$ is also an indicator of entanglement in the reservoirs. The total entropy of $N$ maximally mixed and $N$ pure reservoir qubits is $N$. Therefore, the more that $S_{tot}$ deviates from $N$ as number of iterations $n$ increases, the greater the amount of entanglement. Since entanglement becomes negligible in the weak coupling limit \cite{ziman_quantum_2001}, $S_{tot}$ should remain constant with $n$ for weak coupling, but be greater and vary with $n$ for strong coupling, where the entanglement is significant. 

\begin{figure}[H]
\includegraphics[width=\columnwidth]{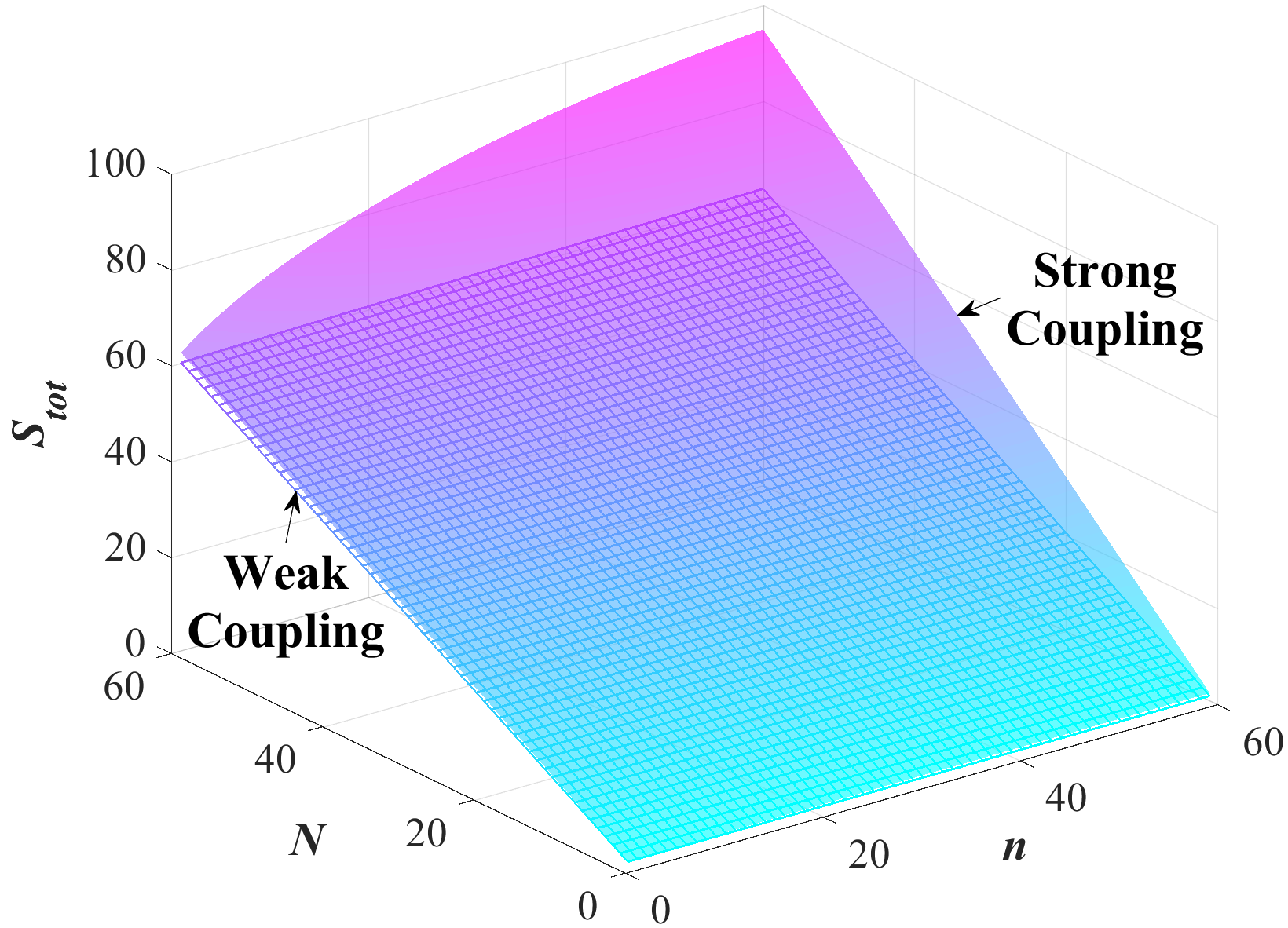}
\caption{\label{total entropies} Total von Neumann entropy of the reservoir qubits in the pure-to-mixed and mixed-to-pure transformation, with $\eta = 0.01$ for weak coupling (flat, meshed surface) and $\eta = 0.1$ for strong coupling (curved, solid surface). $N$ is the number of qubits in each reservoir, and $n$ is the number of iterations of each reservoir.}
\end{figure}

Figure \ref{total entropies} confirms and quantifies this prediction, with the strong coupling surface extending significantly higher in entropy than the weak coupling surface. The difference between the surfaces is a measure of the additional entanglement for strong coupling. This shows that entanglement is significant for strong coupling, but not accounted for in the states used to calculate relative deterioration. This supports our hypothesis in section \ref{RelDet} that the approximation to neglect entanglement has greater impact for strong coupling when $N=n$ is large.

The weak coupling surface remains at a constant $S_{tot}$ with increasing $n$, indicating that the entanglement is negligible, and the changes in von Neumann entropy for the pure-to-mixed and mixed-to-pure tasks are equal and opposite, hence not a cause of irreversibility. 

\subsection{Resources required for homogenization}

\begin{figure*}[!htb]
 \minipage{\columnwidth}
\includegraphics[width=\columnwidth]{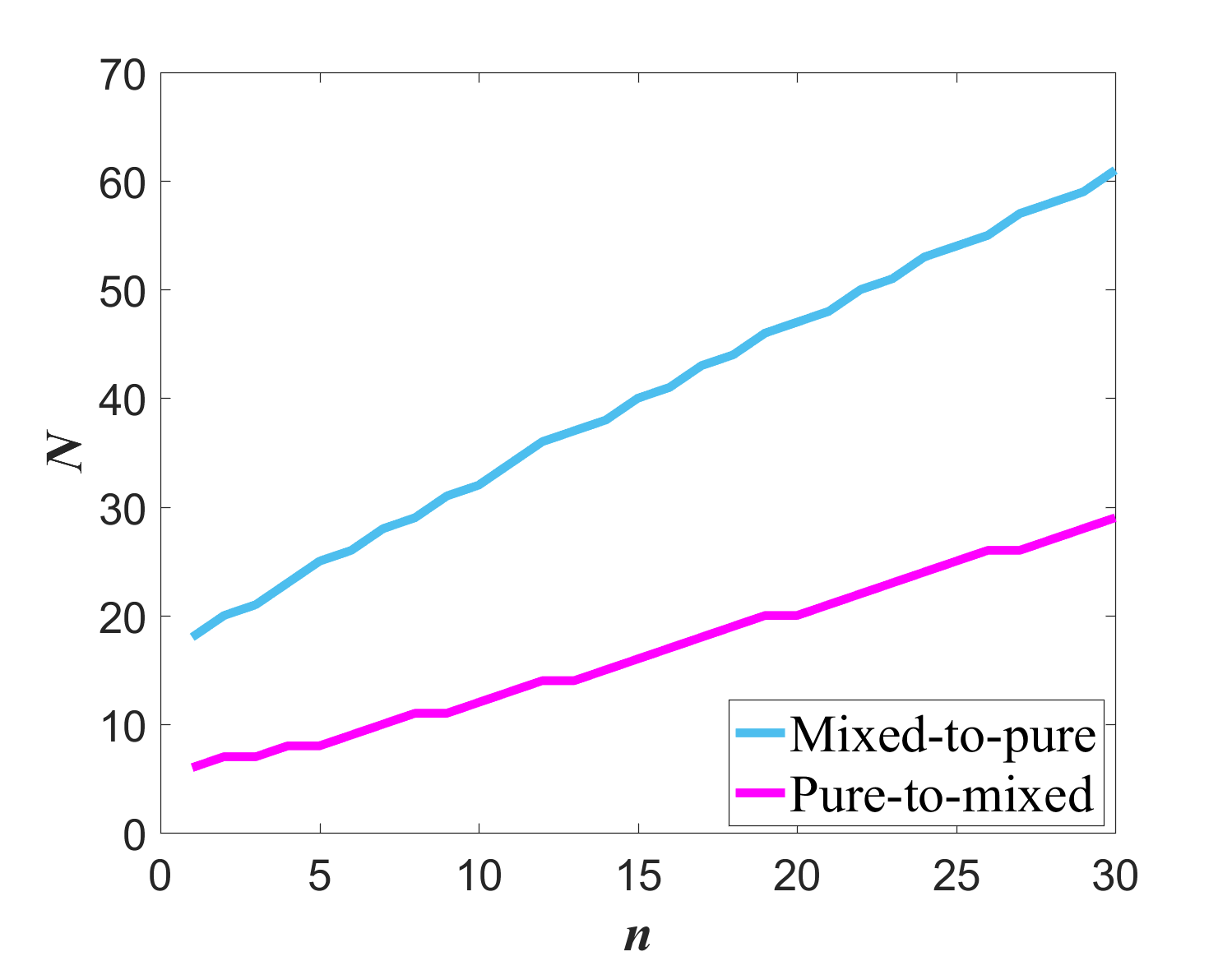}
\caption{\label{resources} Comparison of size of homogenizer required ($N$) to perform a given number of homogenizations ($n$), for a fixed accuracy (error 0.1) and coupling strength (0.3).}
\endminipage\hfill
 \minipage{\columnwidth}
\includegraphics[width=\columnwidth]{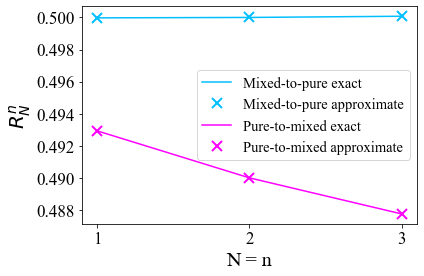}
\caption{\label{rel det 0.01} Comparison of exact (lines) and approximate (crosses) relative deterioration for pure-to-mixed (pink, bottom) and mixed-to-pure (blue, top) transformations for coupling of 0.01, showing good agreement.}
\endminipage
\end{figure*}

Now we will compare the resources required for the pure-to-mixed and mixed-to-pure homogenizations. We can consider the size of the homogenizer needed to achieve a certain number of homogenizations all to a given accuracy, for a fixed coupling strength. Plotting the number of qubits required against number of qubits that can be homogenized to the desired accuracy (number of cycles) we find that the mixed-to-pure transformation requires consistently larger homogenizers than the pure-to-mixed transformation, indicating that the mixed-to-pure transformation requires more resources for repeated use (see Figure \ref{resources}). Furthermore, the mixed-to-pure resources increase more steeply than pure-to-mixed, showing that the increase in resources required become more significant for a greater number of homogenizations. 

Similarly we can invert this relationship to conclude that the lifetime of a mixed-to-pure homogenizer of a given size is consistently lower than that of a pure-to-mixed homogenizer, since it can be used fewer times to achieve the same accuracy. This relationship reinforces the conclusion drawn from the asymmetry in relative deterioration: that the mixed-to-pure transformation is more difficult to perform in a cycle than the reverse process. \\

\subsection{Comparing approximate and exact results}

In the analytic calculation of robustness, we approximated the fidelity of the used reservoir with the original reservoir by treating the entangled reservoir as a product state of its composite qubits (equation \ref{Fidelity product}). We justified this approximation using the fact that the entanglement between reservoir qubits tends to zero for the limit of weak coupling. However, since we use relative deterioration to demonstrate an asymmetry between the mixed-to-pure and pure-to-mixed transformations, one may consider whether the asymmetry could be related to the approximation rather than being a fundamental feature of the transformations. A motivation for this could be that neglecting entanglement seems like a natural way to introduce irreversibility in a system: it amounts to a form of coarse-graining or neglecting information, as is typical for standard derivations of irreversibility. \\

In response to the latter argument, we note that the irreversibility suggested by the asymmetry in relative deterioration is of a different sort to the irreversibility from neglecting information about a system. Constructor-based irreversibility comes from the possibility to perform a transformation arbitrarily well in a cycle, which does not imply that the reverse transformation can be performed arbitrarily well in a cycle; this irreversibility does not emerge by reversing the dynamics of the system. By contrast, neglecting information about a system (or coarse-graining) leads to irreversibility emerging in an approximate or statistical way, such that all evolution of a system is ultimately dynamically reversible if all information about the system is known, and not necessarily reversible if there is neglected information. Therefore, constructor-based irreversibility is not of the type of irreversibility that would naturally emerge from neglecting entanglement in the reservoir. \\

Another reason that we would not expect the approximation to be the cause of asymmetry is that it is used in the same way for both the mixed-to-pure and pure-to-mixed transformations, neglecting entanglement between reservoir qubits in both cases, hence one would expect that it also affects the patterns in relative deterioration in a similar way for both cases.  \\

To furthermore verify the validity of our approximation, we performed exact simulations of the quantum homogenizer and compared these to the approximation. Since the exact simulations become computationally large quickly with number of qubits, we have limited the simulations to maximum of a 3x3 homogenizer, where three system qubits are transformed by a three-qubit reservoir. Specialising to the case where N = n (number of reservoir qubits = number of cycles), we compared the robustness and relative deterioration for the pure-to-mixed and mixed-to-pure transformations, with and without the approximation. We found good agreement between the approximate and exact results for both coupling strengths 0.01 and 0.1. The relative deterioration for coupling 0.01 up to a size 3x3 homogenizer is show in Figure \ref{rel det 0.01}, demonstrating that the predicted asymmetry in relative deterioration holds for an exact calculation and is consistent with the approximation. 

\section{Discussion}

\subsection{Comparison to other approaches}

The notion of re-using quantum machines to perform a task has been studied in a variety of contexts in the field of quantum thermodynamics, for instance through the study of ``catalysts" in the resource theory of thermodynamics \cite{ng_limits_2015, brandao_second_2015}. Our work contrasts to many of these studies in that we do not impose energy conservation or thermal states in the analysis, hence it is a more general model of irreversibility in quantum interactions rather than being a phenomenon specific to thermodynamics. Furthermore, the notion of ``embezzling" has been introduced whereby it appears that using fidelity as a measure of a task's effectiveness is too weak a condition to derive results about irreversibility, when the catalyst has unbounded size \cite{brandao_second_2015}. This contrasts to our discussion of relative deterioration -- this measure shows an irreversibility even when our catalyst has unbounded size and fidelity is used in the distance measure. 

\subsection{Alternative erasers}

We have shown that erasure of a mixed to a pure state using the quantum homogenizer is more difficult than the reverse process. A pertinent question is whether or not this is a special case or a signature of a more general phenomenon, which makes it difficult to generate pure states. One way to assess this would be to consider a different machine that can approximate the task and be approximately re-used, and consider the behaviour of a measure of relative deterioration for such a machine. 

\textbf{Comparison of homogenizer with SWAP.} A candidate for this could be a large reservoir of initially identical qubits, where each interaction of the system with the reservoir is a random swap -- the system qubit is swapped with a randomly selected reservoir qubit. Then as the reservoir becomes large, the system qubit becomes increasingly unlikely to be mistakenly swapped with a previous system qubit rather than one of the original reservoir qubits. There are some key differences between these implementations of erasure. Firstly, with the homogenizer, there is no intrinsic limit on how many times the eraser can be re-used, while still performing erasure to some effectiveness. By contrast the SWAP eraser will either perform perfect erasure or no erasure at all. While the homogenizer deterministically performs erasure with some accuracy, a SWAP eraser would probabilistically perform either perfect or no erasure. It may be that there are certain conditions under which one of the homogenizer and SWAP erasers is more optimal than the other, rather than one being conclusively the best eraser to consider for a task. 

\textbf{Comparison of quantum and classical erasure.} Furthermore we could investigate a toy-model for classical erasure and consider whether the same difficulty of erasure arises. This way we can deduce whether the additional cost of erasure is specific to quantum theory or a more general feature of erasure that works across different regimes. A potential model with which to investigate one form of classical erasure could for instance be through a Controlled-SWAP eraser. Instead of partial swap interactions between system and reservoir qubits, this would instead implement a Controlled-SWAP gate. Each gate is controlled on a qubit in a weighted superposition, where the amplitude of the $\ket{1}$ state varies analogously to the coupling strength in the partial swap, with a system and reservoir qubit as the targets. Such a setup would have fewer quantum effects in play for the homogenization, since there would be no interference and entanglement generated between the reservoir qubits, and between the system qubits. The comparison between the Controlled-SWAP eraser and the quantum homogenizer will be discussed in \cite{beever_comparing_nodate}.

In addition the different erasers can be used as models for different types of physical systems. The homogenizer is closely related to the collision model commonly used to model thermalization processes, with the weak-coupling regime being physically motivated to model systems of interest. By exploring and comparing the limitations and constraints on different forms of erasure, we can reveal how far the limits of erasure vary depending on the environment and physical model being considered. Having said this, there are particular motivations for paying special attention to the quantum homogenizer as an eraser. It is the only machine that can perform erasure and remain arbitrarily close to its initial state \cite{ziman_quantum_2001}. Furthermore there are interesting properties of entanglement that would not emerge with a SWAP or classical eraser, since the information stored in the system qubit becomes spread out across the reservoir qubits in infinitesimal pieces of entanglement, which aids the catalytic behaviour. 

\subsection{Interpretation of the additional cost of erasure}

Our work suggests that there may be an additional cost to erasure when using the quantum homogenizer. While we can conjecture that this cost holds for other forms of erasure, this broader conclusion is not shown by this work. If the cost does not hold for all forms of erasure, then that invites further questions about the space of erasers that are bound by the additional cost. For instance, it may be that the cost appears for quantum but not classical models, or for deterministic but not probabilistic models, or for models with weak coupling between the system and eraser but not for strong coupling. This outcome would also leave open the possibility for other processes or transformations to exhibit constructor-based irreversibility, perhaps connected by a common property of the homogenizer that is different to information erasure. 

Alternatively, if the cost does hold for other forms of erasure, then this suggests a link between constructor-based irreversibility and information erasure. The cost would have practical implications for any implementation of Szilard engines and programmable nanomachines that depend on repeated information erasure. We could then consider whether the irreversibility can be formulated as a constraint on erasure in a more general setting than quantum theory, in terms of CT principles. 

Another natural future development of this work is to better characterise the additional cost of erasure, for instance by expressing the relative deterioration in terms of a type of entropy. We leave these investigations for future work. 

\section{Conclusions}

We have analysed a model of ``constructor-based" irreversibility, which is exact, scale-independent and compatible with the time-reversal symmetric dynamics of quantum theory. We gave a general proof showing how this irreversibility is consistent with quantum theory, and demonstrated an example implementation using the quantum homogenizer. Specifically, the quantum homogenizer cannot be used in a cycle to transform a qubit from a mixed to a pure state, the erasure step of Szilard's engine, but it can be used in a cycle to perform the transpose task. This irreversibility suggests there is an additional cost to the erasure of quantum information, and hence to constructing a reliable implementation of Szilard's engine.  We have quantified how entanglement builds in the homogenizer using von Neumann entropy, and shown that von Neumann entropy is insufficient to capture the additional cost of quantum erasure. We have also demonstrated the difference between the pure-to-mixed and mixed-to-pure tasks through different bounds on the resources required to perform each task in a cycle, with the resources for the latter consistently exceeding those required for the former. Future investigations could examine whether this additional difficulty of generating pure states is specific to the quantum homogenizer or a general feature of erasure. Expressing irreversibility in the way we have demonstrated for erasure could ultimately help us update the status of the second law to be as exact and universal as the conservation of energy.   

\section*{Acknowledgements}

MV is grateful to the Heilbronn Institute for Mathematical Research for their support. This publication was made possible through the support of the ID 61466 grant from the John Templeton Foundation, as part of the The Quantum Information Structure of Spacetime (QISS) Project (qiss.fr). CM’s research was also supported by the Eutopia Foundation and by the grant number (FQXi FFF Grant number FQXi-RFP-1812) from the Foundational Questions Institute and Fetzer Franklin Fund, a donor advised fund of Silicon Valley Community Foundation.

\bibliographystyle{ieeetr}
\bibliography{References3}

\appendix

\section{Recurrence relations} \label{appendix recurrence}

Here we show how the states of system and reservoir qubits are affected by the partial swap operation, leading to recurrence relations for the states. We begin by calculating the effect of the partial swap operation between some reservoir qubit in the state $\xi$ and system qubit $\rho$, with Bloch vectors parallel along some direction $B$: 

\begin{eqnarray} 
\xi = \frac{1}{2} (\mathbb{1} + \alpha \mbox{\boldmath$\sigma \cdot \text{\textbf{B}}$}), \label{res state 1}
\label{Res state}
\\
\rho = \frac{1}{2} (\mathbb{1} + \beta \mbox{\boldmath$\sigma \cdot \text{\textbf{B}}$}). \label{sys state 1}
\label{Sys state}
\end{eqnarray}

The overall resulting state from a unitary operation $U$ acting on the two qubits is given by $U(\rho \otimes \xi)U^\dag$. Here $U$ is the partial swap (equation \ref{partial swap}), which generally entangles the two qubits. We will only be interested in the reduced density operators, defined as: 

\begin{eqnarray}
\xi' = \text{Tr}_1(U(\rho \otimes \xi)U^\dag),
\\
\rho' = \text{Tr}_2(U(\rho \otimes \xi)U^\dag),
\end{eqnarray}

where Tr$_i$ denotes the partial trace over system $i$, here labelling the 1$^{\text{st}}$ or 2$^{\text{nd}}$ state in the tensor product. Using these reduced states in subsequent calculations means that entanglement between the two qubits is neglected. This approximation is dealt with by specialising to the limit of weak coupling, where the entanglement is negligible \cite{ziman_quantum_2001}. 

Computing the partial traces gives the local reservoir qubit state $\xi' = \frac{1}{2} (\mathbb{1} + \alpha' \mbox{\boldmath$\sigma \cdot \text{\textbf{B}}$})$ and system qubit state $\rho' = \frac{1}{2} (\mathbb{1} + \beta' \mbox{\boldmath$\sigma \cdot \text{\textbf{B}}$})$, where

\begin{eqnarray} \label{recurrence states}
\alpha' = \text{s}^2 \beta + \text{c}^2 \alpha, \label{alpha prime}
\\
\beta' = \text{c}^2 \beta + \text{s}^2 \alpha,\label{beta prime}
\end{eqnarray}

 with c $\equiv$ cos$\eta$ and s $\equiv$ sin$\eta$. 

Now we can generalize equations \ref{recurrence states} to arbitrary interactions across the reservoir and over many iterations of the quantum homogenizer. When a reservoir qubit interacts with a new system qubit, the number of iterations $I$ is increased by one (since each reservoir qubit only interacts with a system qubit once per iteration). The index $j$, labelling the position of the reservoir qubit within the entire reservoir of $N$ qubits, stays the same for the interactions of a given reservoir qubit. Hence, if $\beta \equiv \beta^I_j$ then $\beta' = \beta^{I+1}_j$. On the other hand, if $\alpha \equiv \alpha^I_j$ then the system qubit after the partial swap has interacted with the next reservoir qubit in the same iteration, so $j$ is increased by 1 and $I$ stays constant: $\alpha' = \alpha^I_{j+1}$. This means the states for general number of iterations $I$ and reservoir qubit position $j$ can be written: 

\begin{equation} \label{Res recur}
\begin{split}
\alpha^I_j & = \text{s}^2 \beta^I_{j-1} + \text{c}^2 \alpha^{I-1}_j \\
 & = \text{s}^2 \sum_{l=1}^{I}\beta^l_{j-1}+\text{c}^{2I}\alpha^0_I
\end{split}
\end{equation}

and

\begin{equation} \label{Sys recur}
\begin{split}
\beta^I_j & = \text{c}^2 \beta^I_{j-1} + \text{s}^2 \alpha^{I-1}_j \\
& = \text{c}^{2j}\beta^\text{I}_0 + \text{s}^2 \sum_{k=1}^{j}\alpha^{I-1}_{k}
\end{split}
\end{equation}

where we have solved the recurrence relations in terms of the initial system and reservoir states. The equations can be de-coupled by substituting expression \ref{Sys recur} into \ref{Res recur}, such that the recurrence relation for the reservoir qubit state becomes entirely in terms of previous reservoir qubit states. The same can be done for the system qubit recurrence relation by substituting \ref{Res recur} into \ref{Sys recur}. The resulting recurrence relations are: 

\begin{multline}
\beta^I_j = \text{c}^{2j}\Big(\beta^I_0 + ...\\ \text{s}^2\text{c}^{2(I-1)}\sum_{k=1}^{j}\text{c}^{-2k}\Big(\alpha^0_k + \text{s}^2\sum_{l=1}^{I-1}\text{c}^{-2l}\beta^l_{I-1}\Big)\Big),
\end{multline}

\begin{multline}
\alpha^I_j = \text{c}^{2I}\Big(\alpha^0_j + ...\\ \text{s}^2\text{c}^{2(j-1)}\sum_{l=1}^{I}\text{c}^{-2l}\Big(\beta^l_0 + \text{s}^2\sum_{k=1}^{j-1}\text{c}^{-2k}\alpha^{l-1}_{k}\Big)\Big).
\end{multline}

Now the initial conditions for the pure-to-mixed and mixed-to-pure transformations can be substituted into the expressions to derive specific recurrence relations. For the pure-to-mixed case, the initial conditions are a pure system qubit at the start of each iteration ($\beta^I_0 = 1$) and every reservoir qubit being maximally mixed before the first iteration ($\alpha^0_j = 0$). Then the recurrence relations for the pure-to-mixed case are: 

\begin{equation} \label{beta1'}
\beta^I_j = \text{c}^{2j}\Big(1 + \text{s}^4\text{c}^{2(I-1)}\sum_{k=1}^{j}\sum_{l=1}^{I-1}\text{c}^{-2(k+l)}\beta^l_{k-1}\Big),
\end{equation}

\begin{equation} \label{alpha1'}
\alpha^I_j = \text{c}^{2(j-1)}\Big(1 - \text{c}^{2I}+ \text{s}^4\text{c}^{2I}\sum_{l=1}^{I}\sum_{k=1}^{j-1}\text{c}^{-2(l+k)}\alpha^{l-1}_{k}\Big).
\end{equation} \\

Similarly, for the mixed-to-pure case we use the initial conditions $\alpha_0^I = 0$ and $\beta^0_j = 1$ to find: 

\begin{equation} \label{beta2'}
\tilde\beta^I_j = \text{c}^{2(I-1)}\Big(1 - \text{c}^{2j}+ \text{s}^4\text{c}^{2j}\sum_{k=1}^{j}\sum_{l=1}^{I-1}\text{c}^{-2(l+k)}\beta^{l}_{k-1}\Big),
\end{equation}

\begin{equation} \label{alpha2'}
\tilde\alpha^I_j = \text{c}^{2I}\Big(1 + \text{s}^4\text{c}^{2(j-1)}\sum_{l=1}^{I}\sum_{k=1}^{j-1}\text{c}^{-2(l+k)}\beta^{l-1}_{k}\Big).
\end{equation}

\section{Symmetries between states} \label{Appendix Symmetry}

The information to derive all four states is in fact contained within each one, so e.g. equation \ref{alpha1'} is enough to compute the other three. This is due to three independent symmetries constraining the Bloch vectors: 

\begin{eqnarray}
\alpha^I_a + \beta^a_j = 1, \label{Symmetry1}\\
\alpha^a_b = \tilde\beta^b_a,\hspace{0.5cm} \tilde\alpha^a_b = \beta^b_a, \label{Symmetry2}
\end{eqnarray}

where the states with and without a tilde denote the mixed-to-pure and pure-to-mixed tasks respectively. Equation \ref{Symmetry1} can be derived by summing equations \ref{alpha prime} and \ref{beta prime}, with careful attention to the indices, and using the initial condition that the Bloch vector sizes sum to 1 since one of them is 1 and the other is 0. Then equation \ref{Symmetry1} can be derived by induction as follows. 

The sum of Bloch vector sizes for the reservoir qubits on their 0$^{\text{th}}$ iteration and system qubits on their 0$^{\text{th}}$ interaction with the reservoir is always 1: 

\begin{equation}
\alpha^0_j + \beta^I_0 = 1.
\end{equation}

Also, expressing equations \ref{recurrence states} in terms of the indices, and summing them together, gives the general relation: 

\begin{equation} 
\alpha^a_b + \beta^c_d = \alpha^{a+1}_{b} + \beta^{c}_{d+1}.
\end{equation}

Then by induction, one can deduce: 

\begin{equation} 
1 = \alpha^0_j + \beta^I_0 = \alpha^1_j + \beta^I_1 = ... = \alpha^x_j + \beta^I_x.
\end{equation}

For the special case where $j = I$, then: 

\begin{equation} 
\alpha^a_b + \beta^b_a = 1,
\end{equation}

as required. 

Equations \ref{Symmetry2} can be inferred by direct substitution of the indices into the equations \ref{beta1'} to \ref{alpha2'}. They also follow logically when one keeps track of which qubits have interacted with one another. homogenization for a pure system qubit interacting with many mixed reservoir qubits (large $j$) is equivalent to deterioration for a pure reservoir qubit interacting with many mixed system qubits (large $I$).

\section{Error and robustness} \label{Appendix error}

Here we derive the expressions for error, robustness and relative deterioration in terms of the Bloch vector sizes $\alpha$ and $\beta$, providing the expressions shown in table \ref{errors summary}.

\begin{center} 
\begin{tabular}{c||c c} 
\hline \hline
  & Pure to Mixed & Mixed to Pure \\
  \hline \\
 $\epsilon^n_N$ & $\frac{1}{2}(1-\sqrt{1-(\beta^n_N)^2})$ & $\frac{1}{2}(1-\tilde\beta^n_N)$ \\ \\ 
 $\delta^n_N$ & $\prod_{j=1}^{N} \frac{1}{2}(1+\sqrt{1-(\alpha^n_j)^2})$ & $\prod_{j=1}^{N} \frac{1}{2}(1+\tilde\alpha^n_j)$\\ \\
 \hline\hline
\end{tabular}
\captionof{table}{Expressions for error and robustness for the pure-to-mixed and mixed-to-pure tasks.}\label{errors summary}
\end{center}

These quantities are all defined in terms of the fidelity between quantum states, which is a measure of similarity between states. It is 0 for orthogonal states, 1 for identical states, and for two pure states it simply reduces to the inner product. A general result for fidelity of two qubits is \cite{jozsa_fidelity_1994}: 

\begin{equation} \label{Jozsa fidelity}
F(\rho,\xi) =  \text{Tr}(\rho\xi) + 2(\text{det}\rho\text{det}\xi)^{1/2}.
\end{equation}

Using equations \ref{Res state} and \ref{Sys state}, the fidelity can be expressed in terms of Bloch vector sizes:

\begin{equation} \label{Fidelity}
F(\rho,\xi) = \frac{1}{2}\big(1 + \alpha\beta+\sqrt{(1-\alpha^2)(1-\beta^2)}\big).
\end{equation} 

Hence, using equations \ref{beta1'} to \ref{alpha2'}, one can compute the fidelity between any system and reservoir qubits for any values of $I$ and $j$. 

The error is given by: 

\begin{equation}
    \epsilon^n_N = 1 - F(\rho^n_N, \xi).
\end{equation}

The fidelity of a system qubit with the original reservoir qubit can be calculated using equation \ref{Fidelity}. For the pure-to-mixed case, $\alpha^0_j = 0$ giving fidelity:

\begin{equation}
    F = \frac{1}{2}\bigg(1+\sqrt{1-(\beta^n_N)^2}\bigg).
\end{equation}

For the mixed-to-pure case, $\alpha^0_j = 1$ giving fidelity:

\begin{equation}
    F = \frac{1}{2}\bigg(1+\beta^n_N\bigg).
\end{equation}

Subtracting these from 1 to find the error, one deduces that the error for the pure-to-mixed case is: 

\begin{equation}
    \epsilon^n_N = \frac{1}{2}\bigg(1-\sqrt{1-(\beta^n_N)^2}\bigg).
\end{equation}

and for the mixed-to-pure case is: 

\begin{equation}
    \epsilon^n_N= \frac{1}{2}\bigg(1-\beta^n_N\bigg).
\end{equation}

The robustness is given by: 

\begin{equation}
    \delta^n_N = \prod^N_{j=1}F(\xi^0_j,\rho^n_j)
\end{equation}

For the pure-to-mixed case, $\alpha^0_j$ = 0, giving:

\begin{equation}
    F = \frac{1}{2}\bigg(1+\sqrt{1-(\alpha^n_j)^2}\bigg).
\end{equation}

For the mixed-to-pure case, $\alpha^0_j$ = 1, giving:

\begin{equation}
    F = \frac{1}{2}\bigg(1+\alpha^n_j\bigg). 
\end{equation}

Inserting these fidelities into the expression for robustness gives

\begin{equation}
    \delta^n_N = \prod^N_{j=1}\frac{1}{2}\bigg(1+\sqrt{1-(\alpha^n_j)^2}\bigg).
\end{equation}

for the pure-to-mixed case, and 

\begin{equation}
    \delta^n_N = \prod^N_{j=1}\frac{1}{2}\bigg(1+\alpha^n_j\bigg). 
\end{equation}

for the mixed-to-pure case. 

Putting these expressions together, one finds the relative deterioration for the pure-to-mixed case is: 

\begin{equation}
    R^n_N = \frac{\frac{1}{2}\bigg(1-\sqrt{1-(\beta^n_N)^2}\bigg)}{\prod^N_{j=1}\frac{1}{2}\bigg(1+\sqrt{1-(\alpha^n_j)^2}\bigg)}.
\end{equation}

and in the mixed-to-pure case is: 

\begin{equation}
    R^n_N = \frac{\frac{1}{2}\bigg(1-\beta^n_N\bigg)}{\prod^N_{j=1}\frac{1}{2}\bigg(1+\alpha^n_j\bigg)}.
\end{equation}

\end{multicols}


\end{document}